\documentclass[11pt, a4paper]{article}
\usepackage[text={8in,11in},margin=1.0 in,include foot]{geometry}
\usepackage{graphicx}
\usepackage{longtable}
\usepackage{graphicx}
\usepackage{slashed}
\usepackage{amsmath}
\usepackage{ amssymb }
\usepackage{cite}
\begin{document}

\title{Masses and decay widths of radially excited Bottom mesons}
\author{Pallavi Gupta, A. Upadhyay
 \\\small{\it School of Physics and Material Science},\\\small{\it Thapar University,
Patiala, Punjab-147004}\\\small{E-mail: 10gupta.pallavi@gmail.com,
alka.iisc@gmail.com}}

\maketitle

\begin{abstract}
Inspired from the experimental information coming from LHC
\cite{1c,1a} and Babar \cite{1b} for radially higher excited charmed
mesons, we predict the masses and decays of the n=2 S-wave and
P-wave bottom mesons using the effective lagrangian approach. Using
heavy quark effective theory approach,  non-perturbative parameters
($\overline{\Lambda}$, $\lambda_{1}$ and $\lambda_{2}$) are fitted
using the available experimental and theoretical informations on
charm masses. Using heavy quark symmetry and the values of these
fitted parameters, the masses of radially excited even and odd
parity bottom mesons with and without strangness are predicted.
These predicted masses led in constraining the decay widths of these
12 states, and also shed light on the unknown values of the higher
hadronic coupling constants $\widetilde{\widetilde{g}}^{2}_{SH}$ and
$\widetilde{\widetilde{g}}^{2}_{TH}$. Studying the properties like
masses, decays of 2S and 2P states and some hadronic couplings would
help forthcoming experiments to look into these states in future.

\vspace{0.2cm}
 \noindent {\bf PACS: 12.39.Hg, 13.25.Hw, 14.40.Nd}
\vspace{0.2cm}\\
\noindent{\bf Keywords:Heavy quark effective Theory, Decays of
bottom mesons, bottom mesons}
\end{abstract}

\section{Introduction}
In the past decade, many new discoveries have filled the
spectroscopy of charmed and bottom mesons. Recent discovery in 2015,
by LHCb collaboration on bottom states \cite{1} diverts theorists
interest towards the study of bottom sector. Observing the bottom
spectroscopy, it is realized that unlike the success in charm
sector, the experimental information for bottom sector is missing.
Till now experimental information available on bottom mesons is only
for n=1 ground and excited state only which is as shown in Table 1
\cite{2}.

\begin{table}[h]
\begin{center}

\begin{tabular}{|c|c|c|c|c|}

\hline
&\multicolumn{4}{c|}{Experimentally known masses(MeV)}\\
 \cline{2-5}
 $J^{P}(n^{2s+1}L_{J})$&\multicolumn{2}{c|}{Bottom meson}&\multicolumn{2}{c|}{Charm Meson}\\
  \cline{2-5}
&Non-Strange&Strange&Non-Strange&Strange\\
\hline

 $ 0^{-}(1^{1}S_{0})$ & 5279.61/5279.37 &5366.81 &1869.61/1864.84&1968.30\\
$1^{-}(1^{3}S_{1}) $& 5324.83 &5415.4&2010.27/2006.97& 2112.10\\
 $ 0^{-}(2^{1}S_{0})$ & $5840^{*}$\cite{1} &- &$2580^{*}$\cite{1c,1b}&-\\
$1^{-}(2^{3}S_{1}) $&$5960^{*}$\cite{3}  &-&$2680^{*}$\cite{1c,1b}& $2709^{*}$\cite{aa}\\
 $ 0^{+}(1^{3}P_{0})$ & -&-&2403/2318&2317.7 \\
 $ 1^{+}(1^{1}P_{1})$& -&-&2427&2459.5 \\
$ 1^{+}(1^{3}P_{1})$ & 5724.9/5726.8  &5828.4&2421.4&2535.11\\
 $ 2^{+}(1^{3}P_{2})$ & 5739/5735.9 &5839.87&2464.3/2462.6&2571.9 \\
 $ 0^{+}(2^{3}P_{0})$ & -&-&$3000^{*}$\cite{1c,1b}&- \\
 $ 1^{+}(2^{1}P_{1})$&- &-&$3000^{*}$\cite{1c,1b}&$3040^{*} $\cite{ba}\\
$ 1^{+}(2^{3}P_{1})$ & - &-&$3000^{*}$\cite{1c,1b}&$3040^{*}$\cite{ba}\\
 $ 2^{+}(2^{3}P_{2})$ & - &-&$3000^{*}$\cite{1c,1b}/$3214^{*}$\cite{1a}& -\\
$ 1^{-}(1^{3}D_{1})$ & -&-&$2760^{*}$\cite{1c,1b}&$2860^{*}$\cite{ab} \\
 $ 2^{-}(1^{1}D_{2})$& -&-&$2740^{*}$\cite{1c,1b}& -\\
$ 2^{-}(1^{3}D_{2})$ &$ 5840^{*}$\cite{1}&-&$2760^{*}$\cite{1c,1b}&-\\
 $ 3^{-}(1^{3}D_{3})$ &$5960^{*}$\cite{3} &-&$2740^{*}$\cite{1c,1b}&$2860^{*}$\cite{bb} \\

  \hline
\end{tabular}
\caption{Experimentally available Bottom  and Charm Meson Masses.
First value in Columns 2 and Column 4 is for $Q\overline{d}$ and
second value is for $Q\overline{u}$ where Q is the heavy quark i.e.
Q=b/c. Values without $^{*}$ are taken from PDG \cite{2}. States
with $^{*}$ have been experimentally observed, but still their
accurate $J^{P}$ is yet to be assigned. }\label{tab1}

\end{center}
\end{table}

New resonances $B(5970)^{+}$ and $B(5970)^{0}$ are found in the mass
distribution of $B^{0}\pi^{+}$ and $B^{+}\pi^{-}$ respectively.
Masses and decay widths of these resonances as predicted by CDF
collaboration \cite{3} in 2013 are as
\begin{center}
$M(B(5970)^{+})= 5961\pm 13 MeV$

 $\Gamma(B(5970)^{+})= 60\pm MeV $

 $M(B(5970)^{0})=5977\pm13MeV$

 $\Gamma(B(5970)^{0})=70\pm 40 MeV $

\end{center}

Since the resonances decay in $B\pi$ final states, they are expected
to have natural spin-parity states which is still to be confirmed.
Many theorists suggest this state to be $B^{*}(2^{3}S_{1})$
\cite{4,5}. Li-Ye Xio using chiral quark model suggests it to be $1
^{3}D_{3}$ bottom state \cite{6}. And the $B_{J}(5840)$ state
reported by LHCb \cite{1} is suggested to fill the $B(2^{1}S_{0})$
state with $\Gamma(B_{J}(5840)) = $175.9 MeV and $M(B_{J}(5840)) =$
5857 MeV \cite{7}. Many theoretical predictions on the masses and
the decay widths of bottom and charm mesons have been made
\cite{8,9,10,11,12}. These predictions are based on various models
like constituent quark model \cite{8}, pseudo-scalar emission model
\cite{9}, chiral quark model \cite{10}, $^{3}P_{0}$ model \cite{11},
heavy quark effective theory \cite{12} etc. But these different
theoretical predictions are not uniform as these different models
uses different parameters to predict the masses of various states.
As in the non relativistic quark model, Hamiltonian is introduced
which includes various input parameters like r (the separation
between the two quarks), $\sigma$, $\alpha_{s}$, $\mu$, $\gamma_{E}$
etc. As all these are theoretical parameters, so its different input
values generates different mass spectra. And in the framework of
HQET, most of the prediction on masses and the strong decay widths
is available for n=1 states. Information on masses and decays about
the higher states n=2 for strange and non strange sector is not
known clearly due to the presence of extra couplings of the higher
orders. Observing the Table 1, We are motivated to compute the
higher states masses of the spectrum so that we can predict their
decay widths and then can put some constrain on higher hadronic
couplings. In this paper, we made the prediction for masses for n=2
strange and non- strange bottom and charm mesons for H, S and T
fields using the HQET as our model. HQET provides heavy-light meson
mass prediction in terms of few unknown QCD non-perturbative
parameters at a given order of $\frac{1}{m_{Q}}$ \cite{13}. These
parameters $\overline{\Lambda}$, $\lambda_{1}$ and $\lambda_{2}$
represents the operators of the HQET lagrangian at the first order
$\frac{1}{m_{Q}}$ expansion. The information on decays and mass Out
of these parameters, $\lambda_{1}$ gives kinetic energy of the heavy
quark and $\lambda_{2}$ gives the chromomagnetic interaction for the
heavy quark. Previous study on these parameters has provided some
range to their value \cite{14,15} , but this data is least available
for n=2 or higher states. More information about the data for n=2 or
higher states is required to take its value in confidence. Recently
theorists have also predicted the masses of these n=2 states by
using mixing concept. According to this, states with same spin and
parity can mix for e.g. $^{3}P_{1}$ and $^{1}P_{1}$ state can mix as
these states have same spin and parity. Spectroscopy of bottom
mesons attained by this concept, has been shown using models like
non-relativistic quark model \cite{7}, constituent quark model
\cite{16} etc. As it can be seen from the literature that the values
predicted by different models are very much deviating from one
another. Hoping that our calculations provide some insight to our
framework, we proceed as follow: In section 2, a brief review about
the theory used i.e. "Heavy Quark Effective Theory" is given. This
description includes the information about the importance of these
non-perturbative parameters. This is followed by the section 3, in
which  fit these parameters to predict the strange and non-strange
bottom meson masses. These predicted masses are verified by
calculating their strong decay widths in terms of some hadronic
coupling constants followed by the conclusion in the last section.

\section{Framework}

In the framework of the heavy quark effective theory, hadrons
containing single heavy quark are analyzed. This theory is an
effective QCD theory for $N_{f}$ heavy quarks Q with mass $m_{Q}
>>\Lambda_{QCD}$, with heavy quark Q's four velocity fixed \cite{17}. In this
theory, spin and parity of the heavy quark decouples from the light
degrees of freedom quarks as they interact through the exchange of
soft gluons only. Heavy mesons are classified in doublets in
relation to the total conserved angular momentum $i.e.$
$s_{l}=s_{\overline{q}}+l$, where $s_{\overline{q}}$ and
$\textit{l}$ are the spin and orbital angular momentum
 of the light anti-quark respectively. For $\textit{l} =0$ (S-wave) the
doublet is represented by $(P, P^{*})$ with $J^{P}_{s_{l}}=
(0^{-},1^{-})_{\frac{1}{2}}$, which for $\textit{l} =1$ (P-wave),
there are two doublets represented by $(P^{*}_{0},P^{'}_{1})$ and
$(P_{1},P^{*}_{2})$ with $J^{P}_{s_{l}}=(0^{+},1^{+})_{\frac{1}{2}}$
and $(1^{+},2^{+})_{\frac{3}{2}}$ respectively. Two doublets of
$\textit{l}=2$ (D-wave) are represented by $(P^{*}_{1},P_{2})$ and
$(P_{2}^{'},P^{*}_{3})$ belonging to
$J^{P}_{s_{l}}=(1^{-},2^{-})_{\frac{3}{2}}$ and
$(2^{-},3^{-})_{\frac{5}{2}}$ respectively. These doublets are
described by the effective super-field $H_{a}, S_{a}, T_{a}, X_{a},
Y_{a}$ \cite{25}, where the field $H_{a}$ describe the $(P,P^{*})$
doublet i.e. S-wave, $S_{a}$ and $T_{a}$ fields represents the
P-wave doublets $(0^{+},1^{+})_{\frac{1}{2}}$ and
$(1^{+},2^{+})_{\frac{3}{2}}$ respectively. D-wave doublets are
represented by the $X_{a}$ and $Y_{a}$ fields. For the radial
excitation of these states with radial quantum number n=2, these
states are replaced by $\widetilde{P}, \widetilde{P}^{*}$ and so on.
Thus the properties of the hadrons are invariant under $SU(2N_{f})$
transformations, i.e. heavy quark spin and flavor symmetries
providing a clear picture in the study of the heavy quark physics.
These symmetries are exploited to study the charm and bottom meson
spectra and are shown by the QCD lagrangian in the heavy quark
limit. Beyond this symmetry limit, HQET is developed by expanding
the QCD lagrangian in power of $1/m_{Q}$, in which heavy quark
symmetry breaking terms are studied order by order. The QCD
lagrangian for the heavy quark is as:
\begin{gather}
\label{eq:lagrangian} \mathcal{L}_{Q}=\overline{Q}(i
\gamma_{\mu}D^{\mu}-m_{Q})Q
\end{gather}
Where $D^{\mu}\equiv\partial^{\mu}-i gA^{\mu}$. As the interaction
of this heavy quark with light degree of freedom is through the
exchange of soft gluons, which is much smaller than the $m_{Q}$, so
heavy quark momentum $p_{Q}$ is
\begin{center}
\begin{gather}
p^{\mu}_{Q} = m_{Q}v^{\mu}+k^{\mu}
\end{gather}
\end{center}
In this $ m_{Q}v^{\mu}$ is the kinetic momentum which comes from the
mesons's motion and $k^{\mu}$ represents the residual momentum which
is of the order of $\Lambda_{QCD}$. In the $m_{Q} \rightarrow\infty
$ limit, redefining new heavy quark field $h_{v}(x)$, such that it
is related to the original field $Q(x)$ by
\begin{center}
\begin{gather}
\frac{1+\slashed v}{2}Q(x) = e^{-i m_{Q}v.x} h_{v}(x)
\end{gather}
\end{center}
Field $h_{v}(x)$ satisfies
\begin{center}
\begin{gather}
\frac{1+\slashed v}{2}h_{v} = h_{v},\imath \gamma_{\mu}h_{v}(x)=
k^{\mu}h_{v}(x)
\end{gather}
\end{center}
From these relations equation 1 can be reduced to
\begin{gather}
\label{eq:lagrangian} \mathcal{L}_{Q}\rightarrow
\mathcal{L}_{Q,eff}=\overline{h}_{v}(iv.D)h_{v}
\end{gather}
This lagrangian is invariant under both flavor and spin spin
symmetry, since it is independent of heavy quark mass $m_{Q}$ and
the $\overrightarrow{\gamma}$ matrix respectively. Applying finite
heavy quark mass corrections, HQET lagrangian to order of $1/m_{Q}$
is
\begin{gather}
\label{eq:lagrangian}
\mathcal{L}=\overline{h}_{v}(iv.D)h_{v}+\overline{h}_{v}\frac{(iD_{\bot})^{2}}{2m_{Q}}h_{v}+\overline{h}_{v}\frac{g\sigma_{\mu\nu}G^{\mu\nu}}{4m_{Q}}h_{v}+\textrm{O}(\frac{1}{m^{2}_{Q}})
\end{gather}
Where, $D^{\mu}_{\bot}=D^{\mu}-v^{\mu} v.D$ is orthogonal to heavy
quark velocity v, and
$G^{\mu\nu}=T_{a}G_{a}^{\mu\nu}=\frac{\imath}{g_{s}}[D^{\mu},D^{\nu}]$
is the gluon field strength tensor. In the limit
$m_{Q}\rightarrow\infty$, only first term $\overline{h}(iv.D)h$
survives.  This symmetry is broken by the higher order terms in this
lagrangian involving terms of factor $1/m_{Q}$. The second term
$D^{2}_{\bot}$ is arising from the off shell residual momentum of
the heavy quark in the non relativistic model and it represents the
heavy quark kinetic energy $\frac{p^{2}_{Q}}{2m_{Q}}$\cite{18}. This
term breaks the flavor symmetry because of the explicit dependence
on $m_{Q}$, but does not break the spin symmetry of the HQET. The
third term in the above equation i.e. $g\sigma_{\mu\nu}G^{\mu\nu}$
represents the magnetic moment interaction coupling of the heavy
quark spin to the gluon field. This term breaks both the flavor and
spin symmetry. This term is also known as magnetic chromo-magnetic
term. From equation 6, it is seen that heavy quark symmetry is the
symmetry of lowest order of $\mathcal{L}_{Q,eff}$, therefore the
predictions from this heavy quark symmetry are model independent. We
will not consider higher order corrections as we are interested only
upto first order corrections in ($1/m_{Q}$)expansion. Heavy quark
symmetry is used to establish relations between hadron masses. At
$m_{Q}$ order, all hadrons containing same Q are degenerate, i.e.
have the same mass $m_{Q}$ \cite{19}. At the order of unity, the
$\frac{1}{m^{0}_{Q}}$ terms of HQET Hamiltonian ($H_{0}$) obtained
from the first term of lagrangian defined in equation 1 and from the
terms involving light quarks and gluons give contribution to hadron
masses as
\begin{gather}
\frac{1}{2}\langle H^{(Q)}\mid H_{0} \mid H^{(Q)}\rangle \equiv
\overline{\Lambda}
\end{gather}

 At the $1/m_{Q}$ order, there is an extra addition to the hadron
 masses resulting from the contribution coming from the expectation
 value of the $1/m_{Q}$ correction to the Hamiltonian i.e. $H_{1} =
 -L_{1}$.
 Matrix element of two terms in equation 3, define two more
 non-perturbative parameters $\lambda_{1}$ and $\lambda_{2}$ defined
 as:
\begin{gather}
\label{eq:lagrangian} 2\lambda_{1}=-\langle  H^{(Q)}\mid
\overline{h} D^{2}_{\bot} h \mid H^{(Q)}\rangle
\end{gather} and
\begin{gather}
\label{eq:lagrangian}
16(S_{Q}.S_{l})\lambda_{2}(m_{Q})=\alpha(\mu)\langle H^{(Q)}\mid
\overline{h} g \sigma_{\mu\nu}G^{\mu\nu}
 h \mid H^{(Q)}\rangle
\end{gather}
From these two non-perturbative parameters, $\lambda_{1}$ is
independent of $m_{Q}$ and other parameter $\lambda_{2}$ depends on
$m_{Q}$ through the logarithmic $m_{Q}$ dependence of $\alpha(\mu)$
as:
\begin{gather}
\label{eq:lagrangian}
\alpha(\mu)=[\frac{\alpha_{s}(m_{Q})}{\alpha_{s}(\mu)}]^{9/(33-2N_{q})}
\end{gather}
 Sine
$\gamma^{0} h =h$, the matrix element $\overline{h}
\sigma_{\mu\nu}G^{\mu\nu} h$ reduces to the $\overline{h} \sigma . B
h$ , where B is the chromomagnetic field. The operator $\overline{h}
 \sigma h$ represents the heavy quark spin and the matrix component
 of B in the heavy hadron represents the spin of the light degrees
 of freedom. So the contribution to mass from the third term i.e.
the chromomagnetic operator contribution is proportional to
$S_{Q}.S_{l}$. Thus the non-perturbative parameter of this term i.e.
$\lambda_{2}$ transforms like $S_{Q}.S_{l}$ under the spin symmetry.

\subsection{Masses}
 The mass of the
heavy-light hadron to the first order of $1/m_{Q}$ in terms of these
non-perturbative parameters can be represented as :
\begin{gather}
\label{eq:lagrangian}
M_{X}=m_{Q}+\overline{\Lambda}-\frac{\lambda_{1}}{2m_{Q}}+4(S_{Q}.S_{l})\frac{\lambda_{2}}{2m_{Q}}
\end{gather}
In this equation, $d_{H} =-4(S_{Q}.S_{l})$ is the Clebsch factor.
The two parameters $\lambda_{1}$ and $\lambda_{2}$ have the same
value for all the hadrons with same spin-flavor multiplet. The
values of these parameters is of the order of $\Lambda^{2}_{QCD}$.
Since value of kinetic energy of the heavy quark is positive , the
value of the parameter $\lambda_{1}$ should be negative.
$\overline{\Lambda}$ is the HQET parameter whose value is some for
all the particles in a spin-flavor multiplet. The value of
$\overline{\Lambda}_{H}$ for H field mesons is denoted by
$\overline{\Lambda}_{S}$ , for S fields by and for T field by
$\overline{\Lambda}_{T}$ and so on. $\overline{\Lambda}$ does not
depend on the light quark flavor if there is $SU(3)$ symmetry, but
for the breaking of this symmetry $\overline{\Lambda}$ is different
for strange and non-strange heavy -light mesons and is denoted by
$\overline{\Lambda}_{s}$ and $\overline{\Lambda}_{u,d}$
respectively.

Sometimes mass is also written as:
\begin{gather}
\label{eq:lagrangian} M_{X}=m_{Q}+\overline{\Lambda}+\frac{\Delta
m^{2}}{2m_{Q}}+O(\frac{1}{m^{2}_{Q}})
\end{gather}
 Where $\Delta m^{2}$ is related
 to the total spin J of the meson and is given by:
\begin{gather}
\label{eq:lagrangian} \Delta
m^{2}=-\lambda_{1}+2[J(J+1)-\frac{3}{2}]\lambda_{2}
\end{gather}
In these equations X is the hadron in any state, either in ground
state(H) or in excited state(S) or (T),
 $m_{Q}$  is the mass of the heavy quark either c (charm) or b (bottom) making the hadron and J is the total spin of the meson and $\lambda_{1}$, $\lambda_{2}$
are the two non perturbative QCD parameters. $\overline{\Lambda}$
and $\lambda_{1}$ can not be simply estimated by mass measurements
on dimensional grounds. The parameter $\overline{\Lambda}$ gives the
energy of the light degrees of freedom in the limit
$m_{Q}\rightarrow\infty$.

Neglecting $SU(3)$ flavor symmetry breaking, mass relations for the
lowest lying pseudoscalar and vector mesons of $J^{P}=0^{-}$ and
$1^{-}$ respectively i.e. for H fields, $D$ and $D^{*}$ for $Q=c$
and $B$ and $B^{*}$ for $Q=b$ mesons are parameterized as:
\begin{gather}
\label{eq:lagrangian}
m_{H}=m_{Q}+\overline{\Lambda}^{H}-\frac{\lambda_{1}^{H}}{2m_{Q}}-3\frac{\lambda_{2}^{H}}{2m_{Q}}+O(\frac{1}{m^{2}_{Q}})\\
m_{H^{*}}=m_{Q}+\overline{\Lambda}^{H}-\frac{\lambda_{1}^{H}}{2m_{Q}}+\frac{\lambda_{2}^{H}}{2m_{Q}}+O(\frac{1}{m^{2}_{Q}})
\end{gather}
These equations for first orbitally excited state ($l=1$) changes as
shown below. Mass relations for spin $S_{q}=\frac{1}{2}$ i.e. for S
field mesons are
\begin{gather}
\label{eq:lagrangian}
m_{S}=m_{Q}+\overline{\Lambda}^{S}-\frac{\lambda_{1}^{S}}{2m_{Q}}-3\frac{\lambda_{2}^{S}}{2m_{Q}}+O(\frac{1}{m^{2}_{Q}})\\
m_{S^{*}}=m_{Q}+\overline{\Lambda}^{S}-\frac{\lambda_{1}^{S}}{2m_{Q}}+\frac{\lambda_{2}^{S}}{2m_{Q}}+O(\frac{1}{m^{2}_{Q}})
\end{gather}
Similarly for doublet $(1^{+},2^{+})$ belonging to spin
$(S_{q}=\frac{3}{2})$ i.e. for T fields, these relation changes as:
\begin{gather}
\label{eq:lagrangian}
m_{T}=m_{Q}+\overline{\Lambda}^{T}-\frac{\lambda_{1}^{T}}{2m_{Q}}-5\frac{\lambda_{2}^{T}}{2m_{Q}}+O(\frac{1}{m^{2}_{Q}})\\
m_{T^{*}}=m_{Q}+\overline{\Lambda}^{T}-\frac{\lambda_{1}^{T}}{2m_{Q}}+3\frac{\lambda_{2}^{T}}{2m_{Q}}+O(\frac{1}{m^{2}_{Q}})
\end{gather}

These formulas for the difference of spin averaged masses can be
written as:
\begin{gather}
\label{eq:lagrangian}
\overline{m}^{(Q)}_{S}-\overline{m}^{(Q)}_{H}=\overline{\Lambda}^{S}-\overline{\Lambda}^{H}-\frac{\lambda^{S}_{1}}{2m_{Q}}+\frac{\lambda^{H}_{1}}{2m_{Q}}\\
\overline{m}^{(Q)}_{T}-\overline{m}^{(Q)}_{H}=\overline{\Lambda}^{T}-\overline{\Lambda}^{H}-\frac{\lambda^{T}_{1}}{2m_{Q}}+\frac{\lambda^{H}_{1}}{2m_{Q}}
\end{gather}
Where $\overline{m}^{(Q)}_{H}=(3m^{(Q)}_{H^{*}}+m^{(Q)}_{H})/4$,
$\overline{m}^{(Q)}_{S}=(3m^{(Q)}_{S^{*}}+m^{(Q)}_{S})/4$ and
$\overline{m}^{(Q)}_{T}=(5m^{(Q)}_{T^{*}}+3m^{(Q)}_{T})/8$.
Different parameters $\overline{\Lambda}$, $\lambda_{1}$ and
$\lambda_{2}$ appear for different fields. When SU(3) symmetry is
breaking, these parameters are again different for light quarks u, d
and s. Using the above relations and the heavy quark symmetry , some
more relations can be written as \cite{20}:
\begin{gather}
\label{eq:lagrangian}
 \frac{m_{H^{*}}^{b}-m^{b}_{H}}{m_{H^{*}}^{c}-m^{c}_{H}}=\frac{m_{S^{*}}^{b}-m^{b}_{S}}{m_{S^{*}}^{c}-m^{c}_{S}}=\frac{m_{T^{*}}^{b}-m^{b}_{T}}{m_{T^{*}}^{c}-m^{c}_{T}}=\frac{m_{c}}{m_{b}}
\end{gather}
Masses of the heavy hadrons are used to calculate their properties
like strong decays, radiative decays, magnetic moments etc. So
masses can be justified if we know some of the above properties
accurately. In our work, to justify the masses, we study their
strong decay widths.

\subsection{Strong Decays}

Strong interactions are very important for the study of heavy
hadrons containing one heavy and one light quark in the
non-perturbative regime. Heavy meson decay to light pseudo-scalar
meson depends on the initial mass of the heavy hadron and on the
quantum numbers of the decaying resonance. Strong decays are
calculated by approaching heavy meson doublet in effective fields
and imposing the heavy quark spin and flavor symmetry on it
\cite{4}. Strong decay width formulae for $\textit{l}=0,1$ states
decaying to various states are as follow: $(0^{-},1^{-}) \rightarrow
(0^{-},1^{-}) + M$
\begin{gather}
\label{eq:lagrangian} \Gamma(1^{-} \rightarrow 1^{-})=
C_{M}\frac{g_{HH}^{2}M_{f}p_{M}^{3}}{3\pi f_{\pi}^{2}M_{i}}\\
\Gamma(1^{-} \rightarrow 0^{-})=
C_{M}\frac{g_{HH}^{2}M_{f}p_{M}^{3}}{6\pi f_{\pi}^{2}M_{i}}\\
\Gamma(0^{-} \rightarrow 1^{-})=
C_{M}\frac{g_{HH}^{2}M_{f}p_{M}^{3}}{2\pi f_{\pi}^{2}M_{i}}
\end{gather}

 $(0^{+},1^{+}) \rightarrow (0^{-},1^{-}) + M$
\begin{gather}
\label{eq:lagrangian} \Gamma(1^{+} \rightarrow 1^{-})=
C_{M}\frac{g_{SH}^{2}M_{f}(p^{2}_{M}+m^{2}_{M})p_{M}}{2\pi f_{\pi}^{2}M_{i}}\\
\Gamma(0^{+} \rightarrow 0^{-})=
C_{M}\frac{g_{SH}^{2}M_{f}(p^{2}_{M}+m^{2}_{M})p_{M}}{2\pi
f_{\pi}^{2}M_{i}}
\end{gather}

$(0^{-},1^{-}) \rightarrow (0^{+},1^{+}) + M$
\begin{gather}
\label{eq:lagrangian} \Gamma(1^{-} \rightarrow 1^{+})=
C_{M}\frac{g_{SH}^{2}M_{f}(p^{2}_{M}+m^{2}_{M})p_{M}}{2\pi f_{\pi}^{2}M_{i}}\\
\Gamma(0^{-} \rightarrow 0^{+})=
C_{M}\frac{g_{SH}^{2}M_{f}(p^{2}_{M}+m^{2}_{M})p_{M}}{2\pi
f_{\pi}^{2}M_{i}}
\end{gather}

 $(1^{+},2^{+}) \rightarrow (0^{-},1^{-}) + M$
\begin{gather}
\label{eq:lagrangian} \Gamma(2^{+} \rightarrow 1^{-})=
C_{M}\frac{2g_{TH}^{2}M_{f}p_{M}^{5}}{5\pi f_{\pi}^{2}\Lambda^{2}M_{i}}\\
\Gamma(2^{+} \rightarrow 0^{-})=
C_{M}\frac{4g_{TH}^{2}M_{f}p_{M}^{5}}{15\pi f_{\pi}^{2}\Lambda^{2}M_{i}}\\
\Gamma(1^{+} \rightarrow 1^{-})=
C_{M}\frac{2g_{TH}^{2}M_{f}p_{M}^{5}}{3\pi
f_{\pi}^{2}\Lambda^{2}M_{i}}
\end{gather}

$ (0^{-},1^{-}) \rightarrow (1^{+},2^{+}) + M$
\begin{gather}
\label{eq:lagrangian} \Gamma(1^{-} \rightarrow 2^{+})=
C_{M}\frac{2g_{TH}^{2}M_{f}p_{M}^{5}}{3\pi f_{\pi}^{2}\Lambda^{2}M_{i}}\\
\Gamma(1^{-} \rightarrow 1^{+})=
C_{M}\frac{2g_{TH}^{2}M_{f}p_{M}^{5}}{3\pi f_{\pi}^{2}\Lambda^{2}M_{i}}\\
\Gamma(0^{-} \rightarrow 2^{+})=
C_{M}\frac{4g_{TH}^{2}M_{f}p_{M}^{5}}{3\pi
f_{\pi}^{2}\Lambda^{2}M_{i}}
\end{gather}

$ (0^{-},1^{-}) \rightarrow (1^{-},2^{-}) + M$
\begin{gather}
\label{eq:lagrangian} \Gamma(1^{-} \rightarrow 2^{-})=
C_{M}\frac{10g_{HX}^{2}M_{f}(p^{2}_{M}+m^{2}_{M})p_{M}^{3}}{9\pi f_{\pi}^{2}\Lambda^{2}M_{i}}\\
\Gamma(1^{-} \rightarrow 1^{-})=
C_{M}\frac{2g_{HX}^{2}M_{f}(p^{2}_{M}+m^{2}_{M})p_{M}^{3}}{9\pi f_{\pi}^{2}\Lambda^{2}M_{i}}\\
\Gamma(0^{-} \rightarrow 1^{-})=
C_{M}\frac{4g_{HX}^{2}M_{f}(p^{2}_{M}+m^{2}_{M})p_{M}^{3}}{3\pi
f_{\pi}^{2}\Lambda^{2}M_{i}}
\end{gather}

$(0^{+},1^{+}) \rightarrow (1^{-},2^{-}) + M$
\begin{gather}
\label{eq:lagrangian} \Gamma(1^{+} \rightarrow 2^{-})=
C_{M}\frac{2g_{XS}^{2}M_{f}p_{M}^{5}}{3\pi f_{\pi}^{2}\Lambda^{2}M_{i}}\\
\Gamma(1^{+} \rightarrow 1^{-})=
C_{M}\frac{2g_{XS}^{2}M_{f}p_{M}^{5}}{3\pi f_{\pi}^{2}\Lambda^{2}M_{i}}\\
\Gamma(0^{+} \rightarrow 2^{-})=
C_{M}\frac{4g_{XS}^{2}M_{f}p_{M}^{5}}{3\pi
f_{\pi}^{2}\Lambda^{2}M_{i}}
\end{gather}

 $(1^{+},2^{+}) \rightarrow (0^{-},1^{-}) + M$
\begin{gather}
\label{eq:lagrangian} \Gamma(2^{+} \rightarrow 2^{-})=
 C_{M}\frac{17g_{TX}^{2}M_{f}(p^{2}_{M}+m^{2}_{M})p_{M}^{5}}{45\pi f_{\pi}^{2}\Lambda^{2}M_{i}}\\
\Gamma(2^{+} \rightarrow 1^{-})=
 C_{M}\frac{g_{TX}^{2}M_{f}(p^{2}_{M}+m^{2}_{M})p_{M}^{5}}{15\pi f_{\pi}^{2}\Lambda^{2}M_{i}}\\
\Gamma(1^{+} \rightarrow 2^{-})=
 C_{M}\frac{g_{TX}^{2}M_{f}(p^{2}_{M}+m^{2}_{M})p_{M}^{5}}{9\pi f_{\pi}^{2}\Lambda^{2}M_{i}}\\
\Gamma(1^{+} \rightarrow 1^{-})=
 C_{M}\frac{g_{TX}^{2}M_{f}(p^{2}_{M}+m^{2}_{M})p_{M}^{5}}{3\pi f_{\pi}^{2}\Lambda^{2}M_{i}}
\end{gather}

$ (0^{-},1^{-}) \rightarrow (2^{-},3^{-}) + M$
\begin{gather}
\label{eq:lagrangian} \Gamma(1^{-} \rightarrow 3^{-})=
C_{M}\frac{16g_{HY}^{2}M_{f}p_{M}^{7}}{45\pi f_{\pi}^{2}\Lambda^{4}M_{i}}\\
\Gamma(1^{-} \rightarrow 2^{-})=
C_{M}\frac{4g_{HY}^{2}M_{f}p_{M}^{7}}{9\pi f_{\pi}^{2}\Lambda^{4}M_{i}}\\
\Gamma(0^{-} \rightarrow 3^{-})=
C_{M}\frac{4g_{HY}^{2}M_{f}p_{M}^{7}}{5\pi
f_{\pi}^{2}\Lambda^{4}M_{i}}
\end{gather}

$(0^{+},1^{+}) \rightarrow (2^{-},3^{-}) + M$
\begin{gather}
\label{eq:lagrangian} \Gamma(1^{+} \rightarrow 3^{-})=
 C_{M}\frac{28g_{SY}^{2}M_{f}(p^{2}_{M}+m^{2}_{M})p_{M}^{5}}{45\pi f_{\pi}^{2}\Lambda^{4}M_{i}}\\
\Gamma(1^{+} \rightarrow 2^{-})=
 C_{M}\frac{8g_{SY}^{2}M_{f}(p^{2}_{M}+m^{2}_{M})p_{M}^{5}}{45\pi f_{\pi}^{2}\Lambda^{4}M_{i}}\\
\Gamma(0^{+} \rightarrow 2^{-})=
 C_{M}\frac{4g_{SY}^{2}M_{f}(p^{2}_{M}+m^{2}_{M})p_{M}^{5}}{5\pi f_{\pi}^{2}\Lambda^{4}M_{i}}
\end{gather}

$(1^{+},2^{+}) \rightarrow (2^{-},3^{-}) + M$
\begin{gather}
\label{eq:lagrangian} \Gamma(2^{+} \rightarrow 3^{-})=
C_{M}\frac{28g_{TY}^{2}M_{f}p_{M}^{5}}{75\pi f_{\pi}^{2}\Lambda^{2}M_{i}}\\
\Gamma(2^{+} \rightarrow 2^{-})=
C_{M}\frac{7g_{TY}^{2}M_{f}p_{M}^{5}}{75\pi f_{\pi}^{2}\Lambda^{2}M_{i}}\\
\Gamma(1^{+} \rightarrow 3^{-})=
C_{M}\frac{14g_{TY}^{2}M_{f}p_{M}^{5}}{135\pi f_{\pi}^{2}\Lambda^{2}M_{i}}\\
\Gamma(1^{+} \rightarrow 2^{-})=
C_{M}\frac{49g_{TY}^{2}M_{f}p_{M}^{5}}{135\pi
f_{\pi}^{2}\Lambda^{2}M_{i}}
\end{gather}
In the above expressions of decay widths,  $M_{i},M_{f}$ stands for
initial and final meson mass. All hadronic coupling constants are
dependent on the radial quantum number, for n=1 they are notated as
$g_{HH}$, $g_{SH}$ etc, and for coupling between n=2 and n=1 they
will be replaced by $\widetilde{g}^{2}_{HH}$,
$\widetilde{g}^{2}_{SH}$ etc and similarly for the coupling between
initial and final states both belonging to n=2 , they are again
replaced by $\widetilde{\widetilde{g}}^{2}_{HH}$,
$\widetilde{\widetilde{g}}^{2}_{SH}$ etc. These notations can be
made clear from Figure 1. $\Lambda$ is the chiral symmetry breaking
scale $= 1GeV$, $p_{M}$ and $m_{M}$ is the final momentum and mass
of the emitted light pseudo-scalar meson. The coefficient
$C_{\pi^{\pm}}, C_{K^{\pm}}, C_{K^{0}}, C_{\overline{K}^{0}}=1$,
$C_{\pi^{0}}=\frac{1}{2}$ and $C_{\eta}=\frac{2}{3}$ or
$\frac{1}{6}$. Different values of $C_{\eta}$ corresponds to the
initial state being $c\overline{u}, c\overline{d}$ or $
c\overline{s}$ respectively.

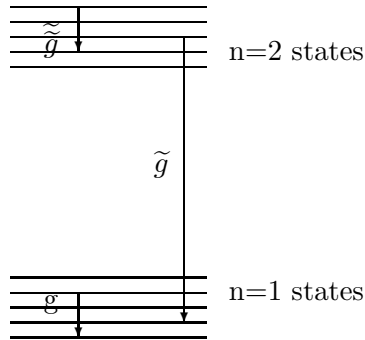
\begin{figure}[h]
\begin{center}
\setlength{\unitlength}{0.20mm}
\begin{picture}(400,250)
\put(75,10){\line(1,0){130}} \put(75,20){\line(1,0){130}}

\put(75,30){\line(1,0){130}} \put(75,40){\line(1,0){130}}
\put(75,50){\line(1,0){130}} \put(75,190){\line(1,0){130}}
\put(75,200){\line(1,0){130}} \put(75,210){\line(1,0){130}}
\put(75,220){\line(1,0){130}} \put(75,230){\line(1,0){130}}
\put(120,230){\vector(0,-1){30}} \put(120,40){\vector(0,-1){30}}
\put(190,210){\vector(0,-1){190}}
\put(97,200){$\widetilde{\widetilde{g}}$}
\put(170,120){$\widetilde{g}$} \put(97,30){g}\put(220,195){n=2
states} \put(220,35){n=1 states}
\end{picture}

\caption{Figure showing the notation for various hadronic couplings}
\label{fig:length_eight_mouse}
\end{center}
\end{figure}


\section{Calculations}
Calculations of this work are subdivided into two parts one in which
we estimate the masses of the bottom mesons and the other in which
the calculated masses are used to predict their strong decay widths.
Decay widths in terms of the hadronic coupling constants are
predicted by constraining the hadronic coupling constants to lie in
the range 0-1.

Mass relations given in equations 14-22 are used for n=2, charm and
bottom mesons for $J^{P}$ $(0^{-},1^{-})$, $(0^{+},1^{+})$ and
$(1^{+},2^{+})$ doublets. These relations are used to fit the values
of the parameters present in these equations
$(\overline{\Lambda}_{}^{H}$, $\lambda_{1}^{H}$, $\lambda_{2}^{H},
\overline{\Lambda}_{}^{S}$, $\lambda_{1}^{S}$, $\lambda_{2}^{S},
\overline{\Lambda}_{}^{T}$, $\lambda_{1}^{T}$ and
$\lambda_{2}^{T})$. Due to the lack of experimental or theoretical
information on these values for radial quantum number n=2, we choose
to fit the differences
$\overline{\Lambda}_{}^{H}-\overline{\Lambda}_{}^{S}$,
$\overline{\Lambda}_{}^{H}-\overline{\Lambda}_{}^{T}$,
$\lambda_{1}^{H}-\lambda_{1}^{S}$ and
$\lambda_{1}^{H}-\lambda_{1}^{T}$, rather than the separate
parameters. As $\lambda_{2}$ parameter gives the mass difference
between the same doublet , so it can be calculated easily once the
masses are known. For this fitting, charm meson masses for n=2 and
the heavy quark symmetry is used. Here heavy quark symmetry implies
that the values of these parameters for charm mesons are same for
the corresponding bottom mesons. Radial charm meson masses for
$J^{P}(0^{-},1^{-})$ doublet are experimentally available as
\cite{21} $\widetilde{D} = 2550$ MeV and $\widetilde{D}^{*}= 2600$
MeV. Masses for two doublets of l=1 (p-wave) are not known
experimentally, but are theoretically predicted by some models
\cite{22,23}. Using the experimental masses of $J^{P}(0^{-},1^{-})$
doublet and taking range of P wave masses as estimated by the
models, we fitted the values of the differences of the parameters as
$\overline{\Lambda}_{}^{H}-\overline{\Lambda}_{}^{S}= 0.29 $ GeV,
$\overline{\Lambda}_{}^{H}-\overline{\Lambda}_{}^{T}= 0.31$ GeV,
$\lambda_{1}^{H}-\lambda_{1}^{S}= 0.28 $ $GeV^{2}$ and
$\lambda_{1}^{H}-\lambda_{1}^{T}= 0.30$ $GeV^{2}$ for the charm
mesons as $\widetilde{D}_{0}^{*}= 2940$ MeV, $\widetilde{D}_{1}^{'}=
3021$ MeV, $\widetilde{D}_{1}= 3031$ MeV and $\widetilde{D}_{2}^{*}=
3032$ MeV. SU(3) breaking gives
$\overline{\Lambda}_{s}^{H}-\overline{\Lambda}_{s}^{S}=0.24$ GeV,
$\overline{\Lambda}_{s}^{H}-\overline{\Lambda}_{s}^{T}=0.28$ GeV,
$\lambda_{1s}^{H}-\lambda_{1s}^{S} = 0.27$ $GeV^{2}$ and
$\lambda_{1s}^{H}-\lambda_{1s}^{T}= 0.30$ $GeV^{2}$ for strange
charm masses $\widetilde{D}_{s} = 2688$ MeV and
$\widetilde{D}_{s}^{*}= 2731$ MeV $\widetilde{D}_{s0}^{*}=3050$ MeV,
$\widetilde{D}_{s1}^{'}=3094$ MeV, $\widetilde{D}_{s1}= 3110$ MeV
and $\widetilde{D}_{s2}^{*}= 3150$ MeV. This fitting is done for
both strange and non-strange radially excited mesons with charm and
bottom quark masses as $m_{c}=1.18$ GeV $m_{b}=4.39$ GeV. This
provides the masses for the bottom mesons which are tabulated in 2
and 5 column of Table 2.

\begin{table}[h]
\begin{center}

\begin{tabular}{|c|c|c|c|c|c|c|}
\hline
&\multicolumn{6}{c|}{Masses of n=2 Bottom Mesons(MeV)}\\
 \cline{2-7}
 $J^{P}(n^{2s+1}L_{J})$&\multicolumn{3}{c|}{Non-Strange}&\multicolumn{3}{c|}{Strange}\\
  \cline{2-7}
&Calculated&\cite{22}&\cite{23}&Calculated&\cite{22}&\cite{23}\\
\hline
$ 0^{-} (2 ^{1}S_{0}) $&5940.64&5890&5886&6022.30&5976&5985\\
  $ 1^{-} (2 ^{3}S_{1})$&5954.04&5906&5920&6033.80&5992&6019\\
  $ 0^{+} (2 ^{3}P_{0})$&6260.84&6221&6163&6301.10&6318&6264\\
  $1^{+} (2 ^{}P_{1})$&6282.61&6281&6194&6313.00&6345&6296\\
  $1^{+} (2 ^{}P_{1})$&6301.10&6209&6175&6340.61&6321&6278\\
  $2^{+} (2 ^{3}P_{2})$&6301.14&6260&6188&6341.14&6359&6292\\
  \hline

  \hline
\end{tabular}
\caption{Theoretically predicted bottom meson masses. Column 2 and 5
represents the non-strange and strange bottom meson masses
calculated in our framework, which are compared with masses
predicted by other theoretical approaches.}\label{tab4}

\end{center}
\end{table}

  These calculated bottom mesons are then used to find calculate the
 $ \lambda_{2}$, which comes out to be  $\lambda_{2}^{H}=29.413$ $GeV^{2}$,
 $\lambda_{2}^{S}=47.76$ $GeV^{2}$ and $\lambda_{2}^{T}=0.087$ $GeV^{2}$. Similarly for the
 strange mesons, this parameter for different fields is calculated as
 $\lambda_{2s}^{H}=25.24$ $GeV^{2}$,
 $\lambda_{2s}^{S}=26.12$ $GeV^{2}$ and $\lambda_{2s}^{T}=1.16$ $GeV^{2}$.

Our calculated masses, are now justified by calculating the strong
decay width for $J^{P}$ $(0^{-},1^{-})$, $(0^{+},1^{+})$ and
$(1^{+},2^{+})$ doublets of bottom meson. Initial masses of these
states are taken from our calculated values mentioned in Table 2,
and the masses of rest of the particles are taken from Ref.
\cite{2,22}. Decay channels along with their decay widths are
presented in Table 3 and in Table 4, for calculated bottom mesons
without and with strangness. Column 3 of these Tables shows the
possible decay channels , whose widths are shown in column 4. In
Column 5, we calculated the total width after using the available
values of hadronic couplings constants in literature \cite{24,25}.

\begin{longtable}{|c|c|c|c|c|}
\hline
 State &  $J^{P}(^{2s+1}L_{J})$ & Decay Channels & Width (MeV)&Total Width (MeV)\\
\hline B (5940.6)&$0^{-} (2 ^{1}S_{0}) $&$B^{*0}\pi^{0}$&778.71$\widetilde{g}^{2}_{HH}$&\\
&&$B^{*+}\pi^{-}$&1579.23$\widetilde{g}^{2}_{HH}$&\\
&&$B^{*0}\eta^{0}$&26.3268$\widetilde{g}^{2}_{HH}$&\\
&&$B^{*}_{s}K^{0}$&34.6502$\widetilde{g}^{2}_{HH}$&189.93\\
&&$B^{*0}_{0}\pi^{0}$&45.5883$\widetilde{g}^{2}_{SH}$&\\
&&$B^{*+}_{0}\pi^{-}$&89.9833$\widetilde{g}^{2}_{SH}$&\\
&&$B^{*+}_{2}\pi^{-}$&0.698535$\widetilde{g}^{2}_{HH}$&\\

\hline $B^{*}$ (5954.0)&$1^{-}(2 ^{3}S_{1})$&$B^{0}\pi^{0}$&337.037$\widetilde{g}^{2}_{HH}$&\\

&&$B^{+}\pi^{-}$&671.146$\widetilde{g}^{2}_{HH}$&\\
&&$B^{0}\eta^{0}$&23.6149$\widetilde{g}^{2}_{HH}$&\\
&&$B_{s}K^{0}$&73.5802$\widetilde{g}^{2}_{HH}$&\\
&&$B^{*0}\pi^{0}$&552.55$\widetilde{g}^{2}_{HH}$&\\
&&$B^{*+}\pi^{-}$&1099.55$\widetilde{g}^{2}_{HH}$&222.44\\
&&$B^{*0}\eta^{0}$&23.2967$\widetilde{g}^{2}_{HH}$&\\
&&$B^{*}_{s}K^{0}$&42.7305$\widetilde{g}^{2}_{HH}$&\\
&&$B^{'0}_{1}\pi^{0}$&32.0787$\widetilde{g}^{2}_{SH}$&\\
&&$B^{'+}_{1}\pi^{-}$&62.6883$\widetilde{g}^{2}_{SH}$&\\
&&$B^{0}_{1}\pi^{0}$&2.4879$\widetilde{g}^{2}_{TH}$&\\
&&$B^{*0}_{2}\pi^{0}$&1.67433$\widetilde{g}^{2}_{TH}$&\\

\hline\hline $B^{*}_{0}(6260.84)$&$0^{+} (2 ^{3}P_{0})$&$B^{0}\pi^{0}$&2918.45$\widetilde{g}^{2}_{SH}$&\\
&&$B^{+}\pi^{-}$&5833.81$\widetilde{g}^{2}_{SH}$&\\
&&$B^{0}\eta$&854.935$\widetilde{g}^{2}_{SH}$&\\
&&$B_{s}K^{0}$&4019.36$\widetilde{g}^{2}_{SH}$&\\
&&$\widetilde{B}^{0}\pi^{0}$&228.942$\widetilde{\widetilde{g}}^{2}_{SH}$&136.26+684.92$\widetilde{\widetilde{g}}^{2}_{SH}$+5.63$\widetilde{g}^{2}_{XS}$+1.10$\widetilde{g}^{2}_{YS}$\\
&&$\widetilde{B}^{+}\pi^{-}$&455.986$\widetilde{\widetilde{g}}^{2}_{SH}$&\\
&&$B_{2}^{0}\pi^{0}$&2.00822$\widetilde{g}^{2}_{XS}$&\\
&&$B_{2}^{+}\pi_{-}$&3.63042$\widetilde{g}^{2}_{XS}$&\\
&&$B_{2}^{'0}\pi^{0}$&0.381716$\widetilde{g}^{2}_{YS}$&\\
&&$B_{2}^{'+}\pi_{-}$&0.722983$\widetilde{g}^{2}_{YS}$&\\
\hline

 $B_{1}$(6282.61) &$1^{+} (2^{}P_{1})$&$\widetilde{B}^{*0}\pi^{0}$&41.1618$\widetilde{\widetilde{g}}^{2}_{SH}$&\\
&&$\widetilde{B}^{*+}\pi^{-}$&80.8277$\widetilde{\widetilde{g}}^{2}_{SH}$&\\
&&$B^{*0}\pi^{0}$&2742.63$\widetilde{g}^{2}_{SH}$&\\
&&$B^{*+}\pi^{-}$&5482.63$\widetilde{g}^{2}_{SH}$&\\
&&$B^{*0}\eta^{0}$&795.318$\widetilde{g}^{2}_{SH}$&\\
&&$B^{*}_{s}K^{0}$&4324.27$\widetilde{g}^{2}_{SH}$&\\
&&$B^{*0}_{1}\pi^{0}$&2.76024$\widetilde{g}^{2}_{XS}$&133.44+121.98$\widetilde{\widetilde{g}}^{2}_{SH}$+13.59$\widetilde{g}^{2}_{XS}$+1.69$\widetilde{g}^{2}_{YS}$\\
&&$B^{*+}_{1}\pi^{-}$&5.16546$\widetilde{g}^{2}_{XS}$&\\
&&$B_{2}^{0}\pi^{0}$&1.98803$\widetilde{g}^{2}_{XS}$&\\
&&$B_{2}^{+}\pi_{-}$&3.68453$\widetilde{g}^{2}_{XS}$&\\
&&$B_{2}^{'0}\pi^{0}$&0.158338$\widetilde{g}^{2}_{SY}$&\\
&&$B_{2}^{'+}\pi_{-}$&0.241994$\widetilde{g}^{2}_{SY}$&\\
&&$B^{*0}_{3}\pi^{0}$&0.444255$\widetilde{g}^{2}_{SY}$&\\
&&$B^{*0}_{3}\pi^{0}$&0.84698$\widetilde{g}^{2}_{SY}$&\\

\hline \hline $B_{1}^{'}$(6301.10) &$1^{+} (2 ^{}P_{1}^{'})$&$\widetilde{B}^{*0}\pi^{0}$&2.00253$\widetilde{\widetilde{g}}^{2}_{TH}$&\\
&&$\widetilde{B}^{*+}\pi^{-}$&3.71221$\widetilde{\widetilde{g}}^{2}_{TH}$&\\
&&$B^{*0}\pi^{0}$&2989.85$\widetilde{g}^{2}_{TH}$&\\
&&$B^{*+}\pi^{-}$&5959.36$\widetilde{g}^{2}_{TH}$&\\
&&$B^{*0}_{1}\pi^{0}$&0.160504$\widetilde{g}^{2}_{TX}$&\\
&&$B^{*+}_{1}\pi^{-}$&0.30404$\widetilde{g}^{2}_{TX}$&289.95+5.71$\widetilde{\widetilde{g}}^{2}_{TH}$+0.57$\widetilde{g}^{2}_{TX}$+19.84$\widetilde{g}^{2}_{TY}$\\
&&$B_{2}^{0}\pi^{0}$&0.0367408$\widetilde{g}^{2}_{TX}$&\\
&&$B_{2}^{+}\pi_{-}$&0.0691139$\widetilde{g}^{2}_{TX}$&\\
&&$B_{2}^{'0}\pi^{0}$&5.45427$\widetilde{g}^{2}_{TY}$&\\
&&$B_{2}^{'+}\pi_{-}$&10.4917$\widetilde{g}^{2}_{TY}$&\\
&&$B^{*0}_{3}\pi^{0}$&1.33718$\widetilde{g}^{2}_{TY}$&\\
&&$B^{*+}_{3}\pi^{-}$&2.56558$\widetilde{g}^{2}_{TY}$&\\

\hline  $B^{*}_{2}$(6301.14)&$2^{+} (2 ^{3}P_{2})$&$\widetilde{B}^{0}\pi^{0}$&22.995$\widetilde{\widetilde{g}}^{2}_{TH}$&\\
&&$\widetilde{B}^{+}\pi^{-}$&45.1337$\widetilde{\widetilde{g}}^{2}_{TH}$&\\
&&$\widetilde{B}^{*0}\pi^{0}$&1.20289$\widetilde{\widetilde{g}}^{2}_{TH}$&\\
&&$\widetilde{B}^{*+}\pi^{-}$&2.22995$\widetilde{\widetilde{g}}^{2}_{TH}$&\\
&&$B^{*0}\pi^{0}$&1794.25$\widetilde{g}^{2}_{TH}$&\\
&&$B^{*+}\pi^{-}$&3576.29$\widetilde{g}^{2}_{TH}$&\\
&&$B^{*0}\eta^{0}$&242.376$\widetilde{g}^{2}_{TH}$&\\
&&$B^{*}_{s}K^{0}$&1344.32$\widetilde{g}^{2}_{TH}$&\\
&&$B^{0}\pi^{0}$&3.02909$\widetilde{g}^{2}_{TH}$&\\
&&$B^{+}\pi^{-}$&3.05293$\widetilde{g}^{2}_{TH}$&254.68+71.56$\widetilde{\widetilde{g}}^{2}_{TH}$+0.45$\widetilde{g}^{2}_{TX}$+18.16$\widetilde{g}^{2}_{TY}$\\
&&$B^{0}\eta$&1.17829$\widetilde{g}^{2}_{TH}$&\\
&&$B_{s}K^{0}$&896.214$\widetilde{g}^{2}_{TH}$&\\
&&$B^{*0}_{1}\pi^{0}$&0.0321397$\widetilde{g}^{2}_{TX}$&\\
&&$B^{*+}_{1}\pi^{-}$&0.0608831$\widetilde{g}^{2}_{TX}$&\\
&&$B_{2}^{0}\pi^{0}$&0.125081$\widetilde{g}^{2}_{TX}$&\\
&&$B_{2}^{+}\pi_{-}$&0.235298$\widetilde{g}^{2}_{TX}$&\\
&&$B_{2}^{'0}\pi^{0}$&1.40358$\widetilde{g}^{2}_{TY}$&\\
&&$B_{2}^{'+}\pi_{-}$&2.69992$\widetilde{g}^{2}_{TY}$&\\
&&$B^{*0}_{3}\pi^{0}$&4.81759$\widetilde{g}^{2}_{TY}$&\\
&&$B^{*0}_{3}\pi^{0}$&9.24343$\widetilde{g}^{2}_{TY}$&\\
  \hline\hline

\caption{Decay widths of calculated masses of non-strange bottom
masses} \label{tab7}

\end{longtable}

\begin{longtable}{|c|c|c|c|c|}
\hline
 State & $J^{P}(^{2s+1}L_{J})$ & Decay Channels & Width(MeV)&Total Width (MeV)\\
\hline
\hline $B_{s}$(6022.5)&$0^{-} (2 ^{1}S_{0}) $&$B^{*0}K^{0}$&807.889$\widetilde{g}^{2}_{HH}$&\\
&&$B^{*+}K^{-}$&827.816$\widetilde{g}^{2}_{HH}$&\\
&&$B^{*}_{s}\pi^{0}$&749.546$\widetilde{g}^{2}_{HH}$&194.32\\
&&$B^{*}_{s}\eta$&84.8866$\widetilde{g}^{2}_{HH}$&\\
&&$B_{s0}^{*}\pi^{0}$&35.6177$\widetilde{g}^{2}_{HS}$&\\
&&$B_{s2}^{*}\pi^{0}$&0.312315$\widetilde{g}^{2}_{HS}$&\\

\hline $B^{*}_{s}$(6033.8)&$1^{-} (2 ^{3}S_{1})$&$B^{0}K^{0}$&410.868$\widetilde{g}^{2}_{HH}$&\\
&&$B^{+}K^{-}$&418.465$\widetilde{g}^{2}_{HH}$&\\
&&$B_{s}\pi^{0}$&328.07$\widetilde{g}^{2}_{HH}$&\\
&&$B_{s}\eta$&86.1936$\widetilde{g}^{2}_{HH}$&\\
&&$B^{*0}K^{0}$&591.59$\widetilde{g}^{2}_{HH}$&\\
&&$B^{*+}K^{-}$&603.842$\widetilde{g}^{2}_{HH}$&239.46\\
&&$B^{*}_{s}\pi^{0}$&527.747$\widetilde{g}^{2}_{HH}$&\\
&&$B^{*}_{s}\eta$&74.789$\widetilde{g}^{2}_{HH}$&\\
&&$B_{s1}^{'}\pi^{0}$&22.1298$\widetilde{g}^{2}_{HS}$&\\
&&$B_{s1}\pi^{0}$&0.503379$\widetilde{g}^{2}_{HS}$&\\
&&$B_{s2}^{*}\pi^{0}$&0.291981$\widetilde{g}^{2}_{HS}$&\\

\hline\hline $B_{s0}^{*}$(6301.1)&$0^{+} (2 ^{3}P_{0})$&$\widetilde{B}_{s}\pi^{0}$&185.906$\widetilde{\widetilde{g}}^{2}_{SH}$&\\
&&$B^{0}K^{0}$&5935.95$\widetilde{g}^{2}_{SH}$&\\
&&$B^{+}K^{-}$&5947.22$\widetilde{g}^{2}_{SH}$&\\
&&$B_{s}\pi^{0}$&2577.88$\widetilde{g}^{2}_{SH}$&174.19+185.90$\widetilde{\widetilde{g}}^{2}_{SH}$+0.18$\widetilde{g}^{2}_{XS}$+0.02$\widetilde{g}^{2}_{YS}$\\
&&$B_{s}\eta$&2958.0$\widetilde{g}^{2}_{SH}$&\\
&&$B_{s2}\pi^{0}$&0.18314$\widetilde{g}^{2}_{SX}$&\\
&&$B_{s2}^{'}\pi^{0}$&0.0256384$\widetilde{g}^{2}_{SY}$&\\

\hline
$B_{s1}$(6313.0)  &$1^{+} (2 ^{}P_{1})$&$\widetilde{B}^{*}_{s}\pi^{0}$&186.491$\widetilde{\widetilde{g}}^{2}_{SH}$&\\
&&$B^{*0}K^{0}$&5400.68$\widetilde{g}^{2}_{SH}$&\\
&&$B^{*+}K^{-}$&5411.62$\widetilde{g}^{2}_{SH}$&\\
&&$B^{*}_{s}\pi^{0}$&2319.87$\widetilde{g}^{2}_{SH}$&\\
&&$B^{*}_{s}\eta$&2609.36$\widetilde{g}^{2}_{SH}$&157.41+186.49$\widetilde{\widetilde{g}}^{2}_{SH}$+0.27$\widetilde{g}^{2}_{XS}$+0.03$\widetilde{g}^{2}_{YS}$\\
&&$B^{*}_{s1}\pi^{0}$&0.194189$\widetilde{g}^{2}_{SX}$&\\
&&$B_{s2}\pi^{0}$&0.0845814$\widetilde{g}^{2}_{SX}$&\\
&&$B_{s2}^{'}\pi^{0}$&0.0101309$\widetilde{g}^{2}_{SY}$&\\
&&$B^{*}_{s3}\pi^{0}$&0.0242715$\widetilde{g}^{2}_{SY}$&\\

\hline \hline $B_{s1}^{'}$(6340.61) &$1^{+} (2 ^{}P_{1}^{'})$&$\widetilde{B}^{*}_{s}\pi^{0}$&34.7237$\widetilde{\widetilde{g}}^{2}_{TH}$&\\
&&$B^{*}_{s}\pi^{0}$&2353.35$\widetilde{g}^{2}_{TH}$&\\
&&$B^{*}_{s1}\pi^{0}$&0.0154948$\widetilde{g}^{2}_{TX}$&\\
&&$B_{s2}\pi^{0}$&0.00266403$\widetilde{g}^{2}_{TX}$&76.24+34.72$\widetilde{\widetilde{g}}^{2}_{TH}$+0.01$\widetilde{g}^{2}_{TX}$+0.24$\widetilde{g}^{2}_{TY}$\\
&&$B_{s2}^{'}\pi^{0}$&1.08377$\widetilde{g}^{2}_{TY}$&\\
&&$B^{*}_{s3}\pi^{0}$&0.2445$\widetilde{g}^{2}_{TY}$&\\

\hline $B_{s2}^{*}$(6341.14)&$2^{+} (2 ^{3}P_{2})$&$\widetilde{B}^{*}_{s}\pi^{0}$&20.9877$\widetilde{\widetilde{g}}^{2}_{TH}$&\\
&&$\widetilde{B}_{s}\pi^{0}$&16.36$\widetilde{\widetilde{g}}^{2}_{TH}$&\\
&&$B^{0}K^{0}$&1949.31$\widetilde{g}^{2}_{TH}$&\\
&&$B^{+}K^{-}$&1971.47$\widetilde{g}^{2}_{TH}$&\\
&&$B_{s}\pi^{0}$&0.552817$\widetilde{g}^{2}_{TH}$&\\
&&$B_{s}\eta$&0.934714$\widetilde{g}^{2}_{TH}$&\\
&&$B^{*0}K^{0}$&2261.54$\widetilde{g}^{2}_{TH}$&342.24+37.34$\widetilde{\widetilde{g}}^{2}_{TH}$+0.01$\widetilde{g}^{2}_{TX}$+1.17$\widetilde{g}^{2}_{TY}$\\
&&$B^{*+}K^{-}$&2289.93$\widetilde{g}^{2}_{TH}$&\\
&&$B^{*}_{s}\pi^{0}$&1415.76$\widetilde{g}^{2}_{TH}$&\\
&&$B^{*}_{s}\eta$&673.684$\widetilde{g}^{2}_{TH}$&\\
&&$B^{*}_{s1}\pi^{0}$&0.00317786$\widetilde{g}^{2}_{TX}$&\\
&&$B_{s2}\pi^{0}$&0.00932343$\widetilde{g}^{2}_{TX}$&\\
&&$B_{s2}^{'}\pi^{0}$&0.282934$\widetilde{g}^{2}_{TY}$&\\
&&$B^{*}_{s3}\pi^{0}$&0.894524$\widetilde{g}^{2}_{TY}$&\\
  \hline\hline

 \caption{Decay widths of calculated masses of
strange bottom masses} \label{tab8}

\end{longtable}

\section{Conclusion}
Last year LHC \cite{1} predicted bottom states, which are assigned
the $J^{P}$ as $1^{+}$ and $2^{+}$ in 1P bottom sector. Experimental
information for the radial excited states 2S, 2P,.. is still
missing. In this paper, we try to shed some light on the masses and
decays of these radial excited 2S and 2P states, by analyzing them
in the heavy quark effective theory. At the $\frac{1}{m_{Q}}$ order,
the bottom meson masses are related to some parameters like
$\overline{\Lambda}$, $\lambda_{1}$ and $\lambda_{2}$. Using the
heavy quark symmetry and the available charm meson masses, we fitted
the $\overline{\Lambda}_{}^{H}-\overline{\Lambda}_{}^{S} $ ,
$\overline{\Lambda}_{}^{H}-\overline{\Lambda}_{}^{T} $ ,
$\lambda_{1}^{H}-\lambda_{1}^{S} $ and
$\lambda_{1}^{H}-\lambda_{1}^{T}$  to attain the masses for the
bottom states. For some best fitted values of these differences, our
predicted masses are comparable with other theoretical models.
Masses calculated in our frame work are about 50 MeV large than the
masses obtained from Ref. \cite{22} and about 126 MeV large than the
values predicted by Ref.\cite{23}. This difference between the
masses, can be reduced by getting a clear information on the unknown
parameters. We assume that values of the differences of these
non-perturbative parameters for n=2 are comparable with their
differences for n=1 states, which is based on that, the difference
of these parameters is independent of the radial number, so that,
they can be used in future to predict the masses of heavy-light
mesons for n=3 quantum number.

Along with the mass prediction, we studied the OZI allowed two body
strong decay to light pseudo-scalar mesons ($\pi, \eta, K$). Column
4 of Table 3 and Table 4 shows the contribution of various possible
decay channels to the total decay width in terms of the various
hadronic coupling constants. As the experimental information about
these hadronic couplings is very limited, so we used the available
theoretical values like $\widetilde{g}^{2}_{HH}$ =0.28 \cite{25},
$\widetilde{g}^{2}_{SH}$=0.1 \cite{24} and
$\widetilde{g}^{2}_{TH}$=0.18\cite{25}, and calculated the total
decay width of these radially excited states in column 5 of these
Tables. $\Gamma(0^{-})$ and $\Gamma(1^{-})$ without strangeness
comes out to be 189 MeV and 222 MeV respectively, and for their
strange partners decay widths comes out to be 194MeV and 239 MeV
respectively.

 As it can be seen from
the column 5 of these Tables, that the contribution to the total
decay width from the decays to X and Y fields is very small, so even
if we vary the values of these couplings $\widetilde{g}^{2}_{SX}$,
$\widetilde{g}^{2}_{SY}$,$\widetilde{g}^{2}_{TX}$ and
$\widetilde{g}^{2}_{TY}$ from 0 to 1, the total decay width would
not effect the result much.

 We are still left
with two more higher hadronic couplings
$\widetilde{\widetilde{g}}^{2}_{SH}$ and
$\widetilde{\widetilde{g}}^{2}_{TH}$. As there is no experimental
information for the decays of these bottom states, the values of
these couplings from 0-1 would effect the total decay width to a
greater extent. To give some insight to these higher order
couplings, we studied the decay widths for higher charm meson
states. As the states $D_{s}(3040)$ and D(3000)and D*(3000) are
expected to fit in 2P charm spectra, so using their experimental
decay widths, we can constrain their couplings to be approximately
$\widetilde{\widetilde{g}}^{2}_{SH}$ =0.1 and
$\widetilde{\widetilde{g}}^{2}_{TH}$=0.3 for the hadronic coupling
$\widetilde{g}^{2}_{SH}$=0.14 and $\widetilde{g}^{2}_{TH}$= 0.12
unlike before.

 So
using $\widetilde{\widetilde{g}}^{2}_{SH}$=0.1,
$\widetilde{\widetilde{g}}^{2}_{TH}$=0.3,
$\widetilde{g}^{2}_{SH}$=0.14 and $\widetilde{g}^{2}_{TH}$= 0.12, we
calculated the total decay width of the n=2 S and P wave bottom
states, which are shown in Table 5. Column 3 of Table 5 shows that
the total decay width deviation for strange states is small due to
the hadronic coupling $g_{SX}$, $g_{SY}$,$g_{TX}$ and $g_{TY}$.

The decay width of T-wave states $B_{1}^{'}$(6228),
$B_{2}^{*}$(6213), $B_{s1}^{'}$(6296)and $B_{s2}^{*}$(6295)comes out
to be $\Gamma(B^{'}_{1})=148 MeV$, $\Gamma(B_{2}^{*})=82 MeV$,
$\Gamma(B_{s1}^{'})=30 MeV$ and $\Gamma(B_{s2}^{*})=90 MeV$
respectively when only decays to light pseudo-scalar mesons are
considered in $^{3}P_{0}$  model \cite{16}. These values are
comparable with $B_{1}^{'}$(6228), $B_{2}^{*}$(6213),
$B_{s1}^{'}$(6296)and $B_{s2}^{*}$(6295) states decay widths
$\Gamma(B^{'}_{1})=139 MeV$, $\Gamma(B_{2}^{*})=128 MeV$,
$\Gamma(B_{s1}^{'})=37 MeV$ and $\Gamma(B_{s2}^{*})=156 MeV$
respectively as calculated in our framework. These decay widths for
non-strange and strange bottom mesons are a kind of motivation for
the theorists and experimentalists to look for them with their
proper $J^{P}$ states to have a clear idea.

\begin{table}[h]
\begin{center}

\begin{tabular}{|c|c|c|}
\hline
&\multicolumn{2}{c|}{Predicted decay Width of n=2 Bottom Mesons(MeV)}\\
 \cline{2-3}
 $J^{P}(n^{2s+1}L_{J})$&Non-Strange&Strange\\

\hline
$ 0^{-} (2 ^{1}S_{0}) $&189.95&194.36\\
  $ 1^{-} (2 ^{3}S_{1})$&223.27&238.67\\
  $ 0^{+} (2 ^{3}P_{0})$&277.29$\pm3.36$&343.37$\pm0.10$\\
  $1^{+} (2 ^{}P_{1})$&270.41$\pm7.64$&310.54$\pm0.15$\\
  $1^{+} (2 ^{}P_{1})$&139.58$\pm10.20$&37.13$\pm0.12$\\
  $2^{+} (2 ^{3}P_{2})$&128.935$\pm9.30$&156.06$\pm0.59$\\
  \hline

  \hline
\end{tabular}
\caption{Theoretically predicted decay width of n=2 bottom meson
masses. Column 2 and 3 represents the non-strange and strange bottom
meson decay widths calculated by us.}\label{tab5}

\end{center}
\end{table}

\section{Acknowledgement}
The authors gratefully acknowledge the financial support by the
Department of Science and Technology (SB/FTP/PS-037/2014), New
Delhi.


\begin{thebibliography}{99}
\bibitem{1}   R. Aaij et al., [LHCb Collaboration], JHEP 1504, 024 (2015) [arXiv:1502.02638 [hep-ex]].
\bibitem{1a}R. Aaij et al, arXiv:1608.01289.
\bibitem{1c} R. Aaij et al., [LHCb Collaboration],JHEP09(2013)145.
\bibitem{1b}P. del Amo Sanchez et al, Phys. Rev. D82 (2010) 111101.
\bibitem{2}  Particle Data Group, K. A. Olive et al., Review of particle physics, Chin. Phys. C38 (2014) 090001.
\bibitem{3}  T. Aaltonen et al. (CDF Collaboration)  Phys. Rev. D 90, 012013 .
\bibitem{aa} B. Aubert et. al., (BABAR Collaboration), Phys. Rev. Lett. 97,222001(2006).
\bibitem{ba} B. Aubert et. al., (BABAR Collaboration), Phys. Rev.D 80, 092003(2009)
\bibitem{ab}R.Aaij, B.Adeva, M.Adinolfi et al., Physical Review
Letters, vol. 113, no. 16, Article ID 162001, 9 pages, 2014.
\bibitem{bb}R.Aaij, B.Adeva, M.Adinolfi et al., Physical Review D, vol. 90, no.
7, Article ID 072003, 2014.
\bibitem{4}  Z.G. Wang ,Eur. phys. J. Plus (2014) 129:186.
\bibitem{5} Hao Xu et. al., Phys. Rev. D89,097502.
\bibitem{6}  Li-Ye Xiao et. al., Phys. Rev. D 90 074029 (2014).
\bibitem{7} Qi-Fang Lu et. al. , arxiv :1607.02812.
\bibitem{8}  S. Godfrey and N. Isgur, Phys. Rev. D 32, 189 (1985)., M. Di Pierro and E. Eichten, Phys. Rev. D 64, 114004
(2001).
\bibitem{9}  S. Godfrey et. al., Phys. Rev. D 89, 074023(2014).
\bibitem{10} L. Y. Xiao and X. H. Zhong, Phys. Rev. D 90, no. 7, 074029 (2014).
\bibitem{11} Y. Sun, Q. T. Song, D. Y. Chen, X. Liu and S. L. Zhu, Phys. Rev. D 89, no. 5, 054026 (2014).
\bibitem{12}Z. G. Wang, Eur. Phys. J. Plus 129, 186 (2014).
\bibitem{13} A. Upadhayay et. al.,  Adv.High Energy Phys. 2014 (2014)619783.
\bibitem{14}. H. D. Politzer and M.B. Wise, Phys. Lett. B 206, 681 (1988).
\bibitem{15} Adam K. Leibovich et. al. Pyhs. Rev. D.  Volume 57, number1, (1997).
\bibitem{16} Stephen Godfrey et. al. , arXiv:1607.02169v2.
\bibitem{17} M. Neubert,Phys. Rept.245(1994)259.
\bibitem{18} Chi-Yee Cheung et. al.,  JHEP04(2014)177.
\bibitem{19} Heavy Quark Physics, Aneesh V. Manohar and Mark B. Wise
(2000).
\bibitem{20} Thomas Mehen et. al., arxiv: 0503134 v2 (2005).
\bibitem{21} P. del Amo Sanchez et al, Phys. Rev. D82 (2010) 111101.
\bibitem{22} D. Ebert, R. N. Faustov and V. O. Galkin, Eur. Phys. J. C66(2010) 197.
\bibitem{23}M. Di Pierro and E. Eichten, Phys. Rev. D64 (2001) 114004.
\bibitem{24} P.Colangelo et. al., Phys.Lett.B642:48-52,(2006).
\bibitem{25} P.Colangelo  et. al., PhysRevD.86.054024 (2012).


\end{thebibliography}
\end{document}